# Vibrational Spectra of MO (M=Sn/Pb) in Their Bulk and Single Layer Forms: Role of Avoided Crossing in their Thermodynamic Properties

Raju K. Biswas,[1] and Swapan K. Pati[1,2,]*

[1]Theoretical Sciences Unit, Jawaharlal Nehru Centre for Advanced Scientific Research (JNCASR), Jakkur P.O., Bangalore 560064, India
[2]School of Advanced Materials and International Centre of Materials Science, Jawaharlal Nehru Centre for Advanced Scientific Research (JNCASR), Jakkur P.O., Bangalore 560064, India.
*E-mail: pati@jncasr.ac.in

## ABSTRACT

We report *ab-initio* calculations of the phonon dispersion relation on the bulk and single layer of SnO and PbO. We identify Raman active modes and infrared active modes at the zone center Γ point. In agreement with experimental observations of Raman spectroscopy measurement, we find that $A_{1g}$ mode is higher in frequency than that of $E_g$ mode. Moreover, the reason behind the shift of $A_{2u}$ mode to higher frequency for monolayer of both SnO and PbO is revealed from our calculations. We also find that long-range Coulomb interaction enhances the dielectric constant and Born effective charges in bulk SnO and bulk PbO, compared to their monolayer. Here, we observe avoided crossing or Landau degeneracy between longitudinal acoustics (LA) and low energetic transverse optical (TO) modes in bulk form of both SnO and PbO. Additionally, monolayer SnO also shows low energetic Raman modes ($E_g$ and $A_{1g}$) of same frequency as bulk. As a result, we notice avoided crossing between LA and TO modes in monolayer SnO. Interestingly, higher Born effective charge and low dielectric constant enhances self-force constants and the interatomic force constants (IFCs) between the M-O bonds. The enhanced force constants give rise to higher vibrational frequency of phonon modes for monolayer PbO. Our studies reveal that due to avoided crossing between two degenerate bands, the phonon dispersion near high symmetry X point lowers specific heat and vibrational entropy in bulk SnO, bulk PbO and only in monolayer SnO upto temperature 150 K. Moreover, the large mass difference between Pb and Oxygen atoms and absence of interlayer van der Waal interactions give rise to high phonon vibration which reduces the occurrence of band crossing between two degenerate energy levels. The absence of avoided crossing leads higher specific heat and vibrational entropy in monolayer PbO at low temperatures.

# INTRODUCTION

Recently, *ab-initio* based density functional theory (DFT) method have become well-established techniques to study vibrational and structural properties of various materials. For the last few decades, many studies on the dynamical property have been carried out on a number of three-dimensional and layered materials. Presently, several mechanical exfoliation techniques, applied to layer materials has lead to the synthesis of a large number of unique two dimensional systems. The most interesting among them is the two-dimensional material, graphene,[1] a hexagonal sheet of carbon atoms, which exhibits exotic physical properties, that are not found in its bulk counterpart, graphite.[2] But the absence of a band gap in graphene makes its application quite limited. On the other hand, there exist a large number of layered crystal structures, like BN-sheets[3], Chalcogenides[4], oxides[5], sulfides[6], etc which show many fascinating properties for applications in electronic, optoelectronic, photovoltaic and thermoelectric devices.

At present, atomic layers of MO (M=Sn/Pb) have been successfully fabricated using micromechanical and sonochemical exfoliation.[7,8] Because of the various band gaps, these materials have attracted attention for their electronic, optoelectronic and thermoelectric applications.[9] Tin monoxide (SnO) is an interesting semiconductor which possesses a relatively small indirect band gap of 0.7 eV and optical gap of 2.7 eV in bulk.[10-12] However, the monolayer is an insulator and theoretically predicted band gap is 3.93 eV at the HSE06 level.[13] The band gap in bulk-SnO can be tuned by reducing the dimension and by inducing biaxial strain effectively.[14] On the other hand, for lead oxide, PbO, the fundamental gap is direct in case of monolayer (4.48 eV) and indirect for both bilayer (3.44 eV) and bulk (2.45 eV).[15]

The major outline of this article is to discuss full phonon spectra of MO (M=Sn/Pb) addressing Raman and IR modes with their associated atomic contributions. We have also performed density functional perturbation theory (DFPT) study to calculate self-force constants and interatomic force constants (IFCs). Interestingly, we obtain negative IFCs between M-M atoms belongs to adjacent layers for both SnO and PbO because of their same charge on each metal atom in two-dimensional layer materials.[16] In the phonon band structure, we encounter avoided crossing between longitudinal acoustic modes and transverse optical modes using the fine q (phonon wave vector) mess calculation. Geurts *et al.* have obtained the vibrational spectrum of single layer and multilayer SnO from Raman spectroscopy measurements.[17] They reported that Raman mode, $A_{1g}$, is higher in frequency than that of $E_g$ mode, which is part of our spectral results. These anomalous behaviour of Raman mode ($A_{1g}$) can successfully be explained by the concept of self-force constants and interatomic force constants. To the best of our knowledge, a fully comprehensive study of vibrational properties of these materials is still absent in the literature.

In Sec. II, we present a computational methodology for the calculation of phonon dispersion and the analysis of force constants. In Sec. III, we discuss the dielectric properties, Born effective charges study and phonon dispersion in two parts for SnO and PbO single layer and bulk and compare with experimental data. In Sec. IV, we calculate self-force constants and interatomic force constants. Landau degeneracy or avoided crossing have been discussed in Sec. V. Finally, the manuscript ends with an overall conclusion in Section VI.

## II. COMPUTATIONAL DETAILS

SnO and PbO are typical layered materials with the same crystal structure as tetragonal symmetry ($D_{4h}$).[18] The tetragonal litharge structure (space group P4/nmm) have 4-coordinate Sn and O atoms in the tetrahedral environment. The Sn atoms are bonded with four neighboring oxygen atoms, which forms the base of the square pyramid.[13,18-20] The asymmetric bonding arrangement gives rise to a layered structure which is stacked along the [001] crystal direction. Here, two consecutive SnO layers are bound by weak van der Waals forces for bulk MO.

The geometry optimization is carried out with density functional theory (DFT) as implemented in the Quantum Espresso package within the plane waves and pseudopotential approach. For structure relaxation and energy calculation, we employ local gradient approximation (LDA) with norm-conserving pseudopotential[21]. The wave function cut-off is considered to be 70 Ry. The Monkhorst-pack grid with 18 × 18 × 18 k-points for bulk MO and 18 × 18 × 1 for monolayer MO are used in sampling Brillouin zone integrations. The crystal structures are fully relaxed to minimize the energy until the magnitude of Hellman-Feynman force on each atom is less than 0.025 eV/ Å. The Gaussian spreading is considered to be 0.003 Ry.

Using first-principle based DFT calculation with van der Waals (vdW) forces, we find the optimized lattice parameters of bulk-SnO as a = b = 3.87 Å and c = 5.02 Å. In the case of PbO, they are a = b = 3.97 Å and c = 5.02 Å. For monolayers, optimized lattice parameters are almost similar to the bulk but there is a vacuum of 20 Å along the crystallographic c-direction. The optimized lattice parameters for both SnO and PbO in their bulk and single layer form are shown in Table I.

Recently, very efficient linear response techniques[22] have been proposed to obtain dynamical matrices at arbitrary wave vectors. Accurate phonon dispersion can be obtained on a fine grid of wave vectors covering the entire Brillouin zone along with the high symmetry points ($\Gamma \rightarrow X \rightarrow M \rightarrow \Gamma$). The optical dielectric constant, the Born effective charge and the force constants matrix have been computed at selected q points in the Brillouin zone using ph.x code implemented in Quantum Espresso. The dynamical matrices has been calculated considering 4x4x4 grid in the Irreducible Brillouin zone (IBZ). The short range contribution to force constants is calculated by considering $4 \times 4 \times 4$ supercell for bulk MO and $4 \times 4 \times 1$ supercell for monolayer MO. Using a discrete Fourier transform, we get the short-range contribution to interatomic force constants (IFCs) in real space. Finally, the phonon dispersion curves can easily be obtained at any arbitrary q-point in the Brillouin zone from the reverse Fourier transform of interatomic force constants in real space.

Since SnO and PbO are polar semiconductors, the long-range character of the Coulomb forces gives rise to the macroscopic electric field. This macroscopic electric field affects the longitudinal optical (LO) phonon modes in the limit of q→ 0, breaking the degeneracy of the LO modes with the transverse optical (TO) modes. In the long wavelength limit, the matrix of the force constants can be written as the sum of analytic and non-analytic contributions of force constants.[23]

$$C_{\alpha i,\beta j} = C^{an}_{\alpha i,\beta j} + C^{na}_{\alpha i,\beta j} \qquad (1)$$

The analytic part, $C^{an}_{\alpha i,\beta j}$, has been calculated from the response of zone-center phonon in zero macroscopic electric field condition. On the other hand, the non-analytic dipole-dipole interaction part of force constant $C^{na}_{\alpha i,\beta j}$, have a general expression

$$C^{na}_{\alpha i,\beta j} = \frac{4\pi e^2}{\Omega} \frac{\sum_{\gamma} Z^*_{i,\gamma\alpha} q_\gamma \sum_{\nu} Z^*_{j,\nu\beta} q_\nu}{\sum_{\gamma,\nu} q_\gamma \varepsilon^\infty_{\gamma\nu} q_\nu} \qquad (2)$$

where $\Omega$ is the volume of the unit cell. $\varepsilon^\infty_{\gamma\nu}$ is the electronic contribution to static dielectric tensor and $Z^*_{i,\gamma\alpha}$ is the Born effective charge tensor for i-th atom in the unit cell. From equation (2), it is clear that non-analytic part of force constants contains macroscopic dielectric constant of the system and Born effective charge, whereas analytic part is calculated ignoring any macroscopic polarization associated with phonons. Once the IFCs are known, we calculate phonon band structure and phonon density of states (DOS) using matdyn.x code implemented in Quantum Espresso.

The computed phonon frequencies ($\omega_{qj}$) calculated over the Brillouin zone can be utilized to enumerate thermodynamic quantity under harmonic approximation.[24] Using thermodynamic relation, thermal properties such as specific heat at constant volume $(C_v)$ and vibrational entropy (S) can be computed as a function of temperature,

$$\text{Specific heat}^{36}, \ C_v = 3nR \int_0^{\omega_{max}} g(\omega) \, E(\frac{\hbar\omega}{k_B T}) \, d\omega \quad (3)$$

$$E(x) = \left(\frac{\frac{x}{2}}{sinh(\frac{x}{2})}\right)^2$$

where n is the number of atoms per formula unit, R is the molar gas constant, $g(\omega)$ is the normalized density of states and $E(x)$ is the Einstein function.

$$\text{Vibrational entropy}^{24}, \ S = \frac{1}{2T} \sum_{qj} \hbar\omega_{qj} coth[\hbar\omega_{qj}/2k_B T] - k_B \sum_{qj} ln[2sinh(\hbar\omega_{qj}/2k_B T)] \quad (4)$$

Where T, $k_B$, $\hbar$ are the temperature, Boltzmann constant and reduced Planck constant.

## III. RESULTS & DISCUSSION

### A. Born effective charge and dielectric properties

To describe long-range dipolar contribution to the lattice dynamics of polar material, it is essential to discuss the Born effective charge ($Z_k^*$) and optical dielectric tensor ($\varepsilon_\infty$)[25]. It is well known from the previous literature[26] study that when Born effective charge correspond to element deviates substantially from its nominal values of oxidation state, the bonding characteristics is no more purely ionic in nature. Our Born effective charge analysis also shows similar results that the associated charge for M and O deviates significantly from its expected oxidation state +2 and −2 respectively, which reflects the partially covalent character of M-O bond both in bulk and monolayer form. It is shown in Table I, the Born effective charge differs more significantly from its ionic oxidation state for bulk SnO than monolayer. As a result, we conclude that Sn-O and Sn-Sn bonds are more covalent nature in bulk compared to its monolayer counterpart. On the other hand, the magnitude of Born effective charges corresponds to Pb and O are higher in monolayer along the xy-plane than the stacking direction. It leads Pb-O bonds in monolayer are more covalent in nature and interlayer Pb-Pb bonds are comparatively weak than interlayer Sn-Sn bonds. Consequently, it has been observed that the Pb-Pb bonds are relatively weak because of the presence of $6s^2$ lone pairs present in the interstitial positions between two layers.[33]

With the optimized geometry, we have calculated electronic contribution to the dielectric tensor self-consistently. The calculated dielectric constant is larger for SnO compared to PbO. The larger value of dielectric constant signifies SnO will allow itself to be more polarized due to the electric field created by their neighbour atomic displacement.[27] From the Born effective charge calculation, we find Sn-O bonds are more ionic in nature compared to Pb-O which gives rise to higher dielectric constant in their bulk and monolayer form of SnO. Weak inter-layer van der Waals interaction increases the dielectric constant along each crystallographic direction in bulk SnO, which is not possible in monolayer (shown in Table I). Our density functional perturbation theory (DFPT) calculations estimated the dielectric constant to be 7.47 along the xy-direction and 6.86 along z-direction for bulk SnO, whereas the dielectric constant is 2.18 along the xy-direction and 1.32 along z-direction for monolayer SnO. Moreover, we have computed dielectric constant and Born effective charge for PbO that exhibits a similar trend like SnO (see Table I). This large difference in values for bulk and monolayer is an indicator that weak inter-layer van der Waals interactions play an important role to increase dielectric tensor and $Z_k^*$ in both bulk SnO and PbO layered materials.

## B. Phonon Dispersion

### 1. SnO

We begin our analysis of the vibrational properties with the description of the general features of the phonon dispersion of bulk and single layer MO. We have compared our theoretical results with the experimental available data.[19,28-30] The overall agreement between theoretical lattice parameters and vibrational modes with experimental results motivates us to choose LDA functional for our further calculations (see Table I,II,III). We have also verified theoretically predicted vibrational modes of bulk form of both SnO and PbO with the experimental results.[17,34,35]

The tetragonal unit cell of SnO consists of 4 atoms, so there are 12 normal modes of vibration (three acoustics modes and nine optical modes) at any q point in the Brillouin zone. At the zone center ($\Gamma$), three acoustics (ac) modes with $\Gamma_{ac} = 2E_u + A_{2u}$ symmetry have zero frequencies. The optical (op) modes are ( $\Gamma_{op} = 4E_g + A_{1g} + 2E_u + B_{1g} + A_{2u}$) for bulk-SnO. Here g and u subscript represent symmetric and antisymmetric modes with respect to the center of inversion. The calculated eigenvectors allow us to visualize every mode in terms of motions of Sn and O atoms. The modes which vibrate in-plane [longitudinal acoustic (LA) and transverse acoustic (TA)] have dispersion higher than that of out of plane mode (ZA) which is shown in Figure 2. There are low energy optical modes found at 104 cm$^{-1}$ and 199 cm$^{-1}$ . As phonon wave vector (q) increases, longitudinal acoustic modes (LA) experiences avoided scattering with the transverse optical modes (TO) along the high symmetry path $\Gamma$ to X (T-point), which will be discussed later.

The high energy optical modes are separated from low energy optical modes by the gap of 65 cm$^{-1}$ at $\Gamma$ point. We have also identified the atomic displacement of Raman active modes $E_g$, $A_{1g}$, $B_{1g}$ and infrared active modes $E_u$ and $A_{2u}$ by studying the eigenvectors of each mode using XCrySDen programme[31]. The vibrational eigenvectors corresponding to modes $A_{1g}$, $B_{1g}$, $A_{2u}$ are shown in Figure S1 (Supplementary material). The Raman and infrared (IR) modes are also tabulated in Table II. From the Table II, we observe that the low energy Raman modes ($E_g$, $A_{1g}$) corresponding to bulk SnO are same as monolayer SnO, but this is not the case for PbO. This anomaly in the Raman modes will be discussed with the help of self-force constants and IFCs. We also notice that IR mode ($A_{2u}$) is lower in frequency than $E_g$ Raman modes in bulk-SnO (see Table II). But on the Contrary, we find $A_{2u}$ mode is higher in frequency than $E_g$ for monolayer SnO which is because of lower dielectric constant (1.32) along the stacking directions. The lower dielectric constant increases the interatomic force constants (see equation 2) and in turn increases mode vibration. Moreover, since we neglect the weak interlayer interaction in monolayer SnO, one IR mode, $A_{1u}$, shifts to a higher frequency (from 383.36 cm$^{-1}$ to 506.95 cm$^{-1}$).

The vibrational modes, particularly, low energy modes are shown in the dispersion curve for bulk and monolayer SnO (see Figure 2a and 2b). There are 6 Raman modes tabulated along with atomic contribution in Table II. There are also another 6 IR active modes including three acoustics modes. Frequencies of the first three Raman modes are almost the same for bulk and monolayer SnO (104 cm$^{-1}$, 104 cm$^{-1}$, 199 cm$^{-1}$). Reasons behind this will be discussed later in the IFCs part. Furthermore, phonon modes corresponding to monolayer are higher in magnitude with respect to its bulk counterpart that is due to absence of interlayer vdW interactions. Overall, monolayer and bulk SnO phonon dispersion show similar behavior. Nevertheless, we analyze all the above features in the next section with the help of self-force constant and the interatomic force constants (IFCs).

## 2. PbO

Figure 3 shows the phonon dispersion curve of both bulk and single layer PbO. The general features are almost identical with SnO phonon dispersion curve. For a better comparison of SnO and PbO phonon bands in single layer and bulk, we have plotted phonon band structures in figure 2 and 3 respectively. We observe that $A_{1g}$ modes is lower in frequency for monolayer PbO than bulk-PbO. This peculiar feature can be understood using the concept of self-force constants and IFCs. The larger mass difference between Pb and Oxygen leads to higher frequency gap between low and high-frequency optical modes. The frequency gap between low and high energy optical modes for bulk and a single layer of SnO are around 66 cm$^{-1}$ and 76 cm$^{-1}$ respectively whereas for bulk and single layer PbO, this gaps are around 105 cm$^{-1}$ and 227 cm$^{-1}$ respectively. The larger frequency gap for single layer PbO is the consequence of higher mass difference between Oxygen and Pb. Moreover, the frequency gap between low and high energy optical mode for monolayer PbO is further larger than bulk counterpart. This could be because of absence of interlayer van der Waal interactions. All vibrational modes for PbO are mentioned in Table III. Moreover, vibrational eigenvector corresponding to modes $A_{1g}$, $B_{1g}$, $A_{2u}$ are shown in Figure S2 (Supplementary material)

## IV. INTERATOMIC FORCE CONSTANT

The interatomic force constants (IFCs) are considered as a fitting parameter in the vibrational problem of nonmetallic crystals. It is merely an expansion coefficient of an adiabatic potential with respect to the atomic displacements. The interatomic force constants (IFCs) are calculated in the construction of phonon dispersion relation. We consider that the IFCs matrix

$C_{(\alpha,\beta)}(lk, l'k')$ which relates the force $F_\alpha(lk)$ on an atom k of unit cell $l$ due to displacement $\Delta\tau_\beta(l'k')$ of the atom $k'$ in the cell $l'$ is defined through the following expression,

$$F_\alpha(lk) = C_{(\alpha,\beta)}(lk, l'k') \, \Delta\tau_\beta(l'k')$$

Initially, we examine self-force constants which specifies the force on a single isolated atom at a unit displacement from its crystalline position. The values are tabulated in Table IV. The self-force constants are positive for all atoms in the two compounds both in bulk and monolayers. The positive value in self-force constant suggests that these two compounds are stable against isolated atomic displacement in their bulk and monolayer stable form.

We observe that there are three low energetic Raman modes (104 cm$^{-1}$, 104 cm$^{-1}$ and 199 cm$^{-1}$) which are almost the same for both bulk and monolayer SnO. The eigenvectors corresponding to the Raman modes $E_g$ is mainly contributed by both Sn and O atomic displacement and the other Raman mode $A_{1g}$ is originated by solely Sn displacement (See Figure S1). Since self-force constant of Sn and O are the same for both bulk and monolayer SnO, $E_g$ vibrations are also same. On the other hand, interatomic force constant between interlayer Sn atoms has negative value which leads to increase in frequency (see Table V). As a result, $A_{1g}$ mode has higher frequency (199.94 cm$^{-1}$) in bulk SnO compared to its monolayer counterpart (199.06 cm$^{-1}$). Two IR modes $E_u$ are also same in frequency (265.46 cm$^{-1}$) and they are associated with Sn and O oscillations. One IR mode $A_{2u}$ and two other Raman modes $E_g$ are mainly associated with Oxygen oscillations. Since O atoms are lighter in mass, these modes ($A_{2u}$, $E_g$) show higher vibrational frequencies.

Now we discuss phonon vibrational spectra through the analysis of real space force constants for monolayer SnO. In this case, one IR mode $A_{2u}$, shifts higher in frequency for monolayer SnO which is because of the absence of interlayer van der Waals interaction along the z-direction (see Table II). Another reason could be due to higher value of self-force constant along the stacking directions for monolayer with respect to bulk SnO. The self-force constants of Sn atom are more pronounced along out of plane direction which leads to the higher frequency of mode $A_{1u}$ with respect to $E_g$ modes (see Table IV).

Unlike SnO, we observe that the first two Raman mode $E_g$ are different in magnitude for bulk and monolayer PbO. The reason is that the self-force constant for Pb is higher for monolayer with respect to bulk in the in-plane direction (see Table IV). The three Raman mode ($B_{1g}$, $2E_g$) and three IR modes ($2E_u$, $A_{2u}$) are higher in frequency for monolayer compared to its bulk form. This is not only because of large self-force constant calculated using equation 2 but also the absence of interlayer interactions which give rise to higher vibrational frequencies. The eigenvectors corresponding to $A_{1g}$, $B_{1g}$, $A_{2u}$ are shown in Figure S2 (supplementary material).

## V. AVOIDED CROSSING

Avoided crossing or Landau degeneracy occurs where two closely spaced energy levels with the same symmetry do not cross each other.[32] On the other hand, energy levels with different symmetry may cross each other at any q point in the phonon band structure. The gap between these two avoided bands depends upon the interaction strength in the vicinity of the avoided crossing point. Our calculations and subsequent analysis reveals that there is Landau degeneracy or avoided crossing due to the coupling between low lying optical (LLO) mode and longitudinal acoustic (LA) mode along the path Γ to M direction (Σ point) and Γ to K directions (T point) in the phonon dispersion curve (See Figure 2 and 3). We observe avoided crossing in a single layer and bulk structure of SnO and only in bulk PbO. Interestingly, we could not find any band crossing near T and Σ points in the phonon dispersion curve of monolayer PbO.

Figure 4 implies avoided crossing of two degenerate bands in the phonon dispersion band due to strong hybridization between the transverse optical (TO) and longitudinal acoustics (LA) modes. These hybridized vibrational modes are identified using the point group symmetry of each individual mode. To analyze the band crossing of the Landau degenerate bands in the vicinity of the avoided crossing point, the phonon eigenvectors are determined by diagonalizing the dynamical matrices. These phonon eigenvectors shown in Figure 4 indicates both phonon modes at the band crossing point have same point group symmetry ($E_u$) and the magnitude of the eigenvectors signify equal contribution from TO and LA modes at particular q point (T & Σ points), where the avoided crossing occurs. Avoided crossing is further verified from the phonon density of states calculations shown in Figure 5. Since avoided crossing is absent in monolayer PbO, we observe finite density of states within the temperature range between 100 K to 150 K. Due to band crossing between two degenerate bands in bulk and monolayer SnO and only in bulk PbO in the frequency range 90 $cm^{-1}$ to 120 $cm^{-1}$, we find less density of states up to 150 $cm^{-1}$ (see Figure 5).

As can be seen in Figure 4, avoided crossing occurs at a frequency from 90 $cm^{-1}$ to 91 $cm^{-1}$. Note that, this frequency range is equivalent to temperature, ≈ 150 K. As a consequence of avoided crossing in the phonon band structure, we observe the significant amount of effect on thermodynamic quantities. Figure 6 illustrates the specific heat and vibrational entropy at various temperatures from O K to 700 K, calculated using equation 4 and 5 respectively. We have shown that the specific heat and vibrational entropy are lower for both bulk and monolayer SnO and only in bulk PbO than for monolayer PbO, up to T ≈ 150K (see Figure 6). It is because of the avoided crossing, which is observed in both bulk and monolayer SnO along with bulk PbO. Interestingly, we could not notice any avoided crossing for single layer PbO. This is due to the presence of higher vibrational mode $E_g$ of 108.25 $cm^{-1}$ in monolayer PbO in comparison to the same mode in bulk PbO at 81.79 $cm^{-1}$. However, in both bulk and monolayer SnO, this vibrational mode, $E_g$, has the same frequency of 104.39 $cm^{-1}$.

At higher temperature, the specific heat is lower for PbO than for SnO due to the heavier mass of Pb compared to Sn. Heavier mass decreases the phonon frequency and as a result, specific heat and vibrational entropy also decrease at the higher temperatures. Our temperature dependent thermodynamic parameter study also shows that there is avoided crossing near the frequency, 91 $cm^{-1}$, for SnO and bulk PbO, which is manifested in lower specific heat and entropy till T ≈ 150 K.

To further confirm the avoided crossing between two degenerate bands in bulk SnO, bulk PbO and monolayer SnO, we plot $C_v/T^3$ at various temperatures (see Figure 7). Here, we have shown that $C_v/T^3$ value is higher for monolayer PbO than the bulk form of SnO and PbO along with monolayer SnO. The lower value of $C_v/T^3$ further confirms that the appearance of avoided crossing in the dispersion relation between two degenerate bands upto temperature around 150 K.

## VI. CONCLUSION

In conclusion, we have compared and contrasted phonon dispersions of SnO and PbO in their single layer and bulk forms using Density Functional Perturbation Theory (DFPT). We obtain good agreement of Raman and IR active modes between our theory and the available experimental data. We have shown some anomalous behavior of phonon dispersions which can be understood from the study of self-force constants and IFCs. Interestingly, we notice strong avoided crossing in both monolayer and bulk SnO and only in bulk PbO along the high symmetry points, which has a direct consequence in reducing phonon specific heat and vibrational entropy. Our $C_v/T^3$ plot further confirm the appearance of Landau degeneracy or avoided crossing in bulk and monolayer form of SnO and only in bulk PbO. Interestingly, we do not find any avoided crossing only for monolayer PbO which is additionally validated by $C_v/T^3$ plot in Figure 7.

## ACKNOWLEDGMENTS

Authors thank Dr. Y. Anusooya Pati for helpful discussions. RKB thanks the UGC, Govt. of India and SKP acknowledges JCBose Fellowship, SERB, DST, Govt. of India, for financial support.

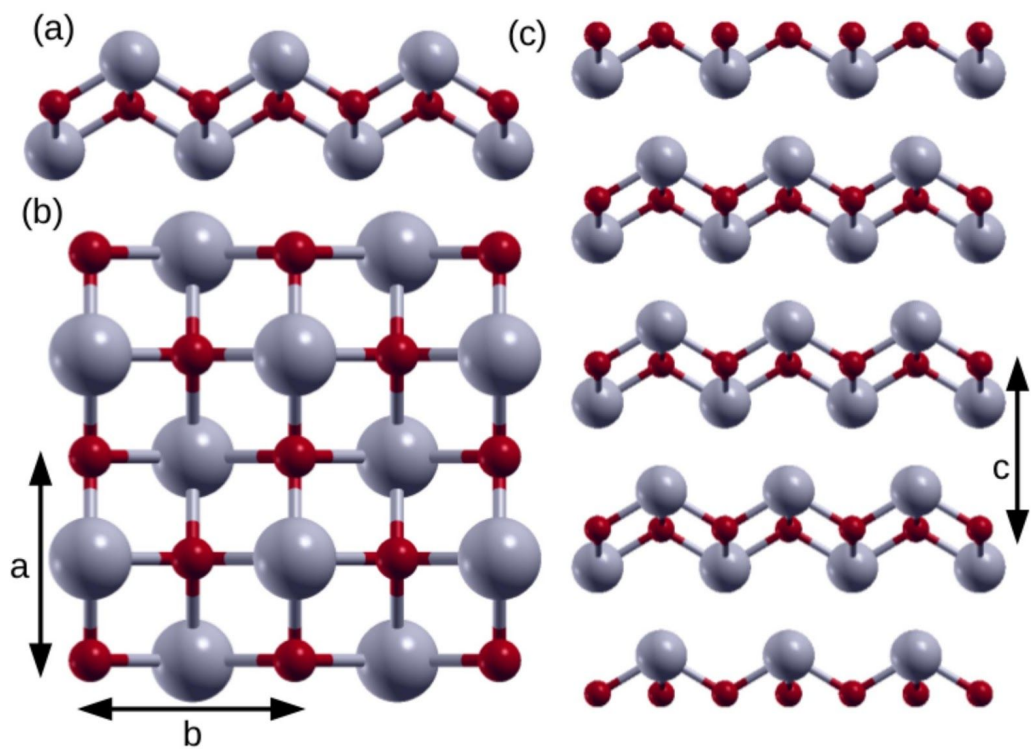

**Figure 1:** Crystal structure of MO (M=Sn/Pb). (a) monolayer (b) viewed along z-axis, (c) bulk. The arrows in (b) depict lattice parameters a and b along the x and y directions, respectively. The interlayer distance is shown by arrow in (c).

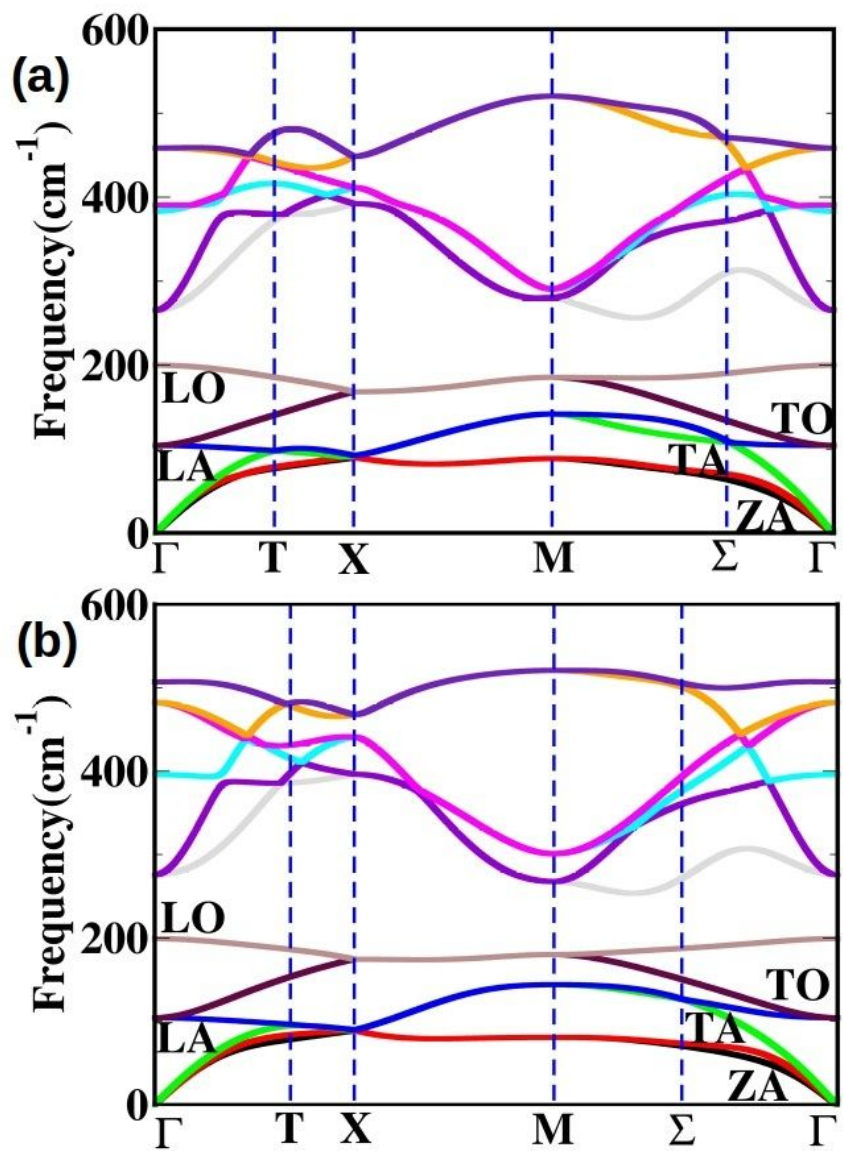

**Figure 2:** Phonon band structure of (a) bulk and (b) monolayer of SnO. Low energy acoustics and optical modes are indicated in the graph using several colours. The bands are represented by different colours like black: ZA, red: TA, green: LA, blue,maroon: TO (transverse optical), gray: LO (longitudinal optical).

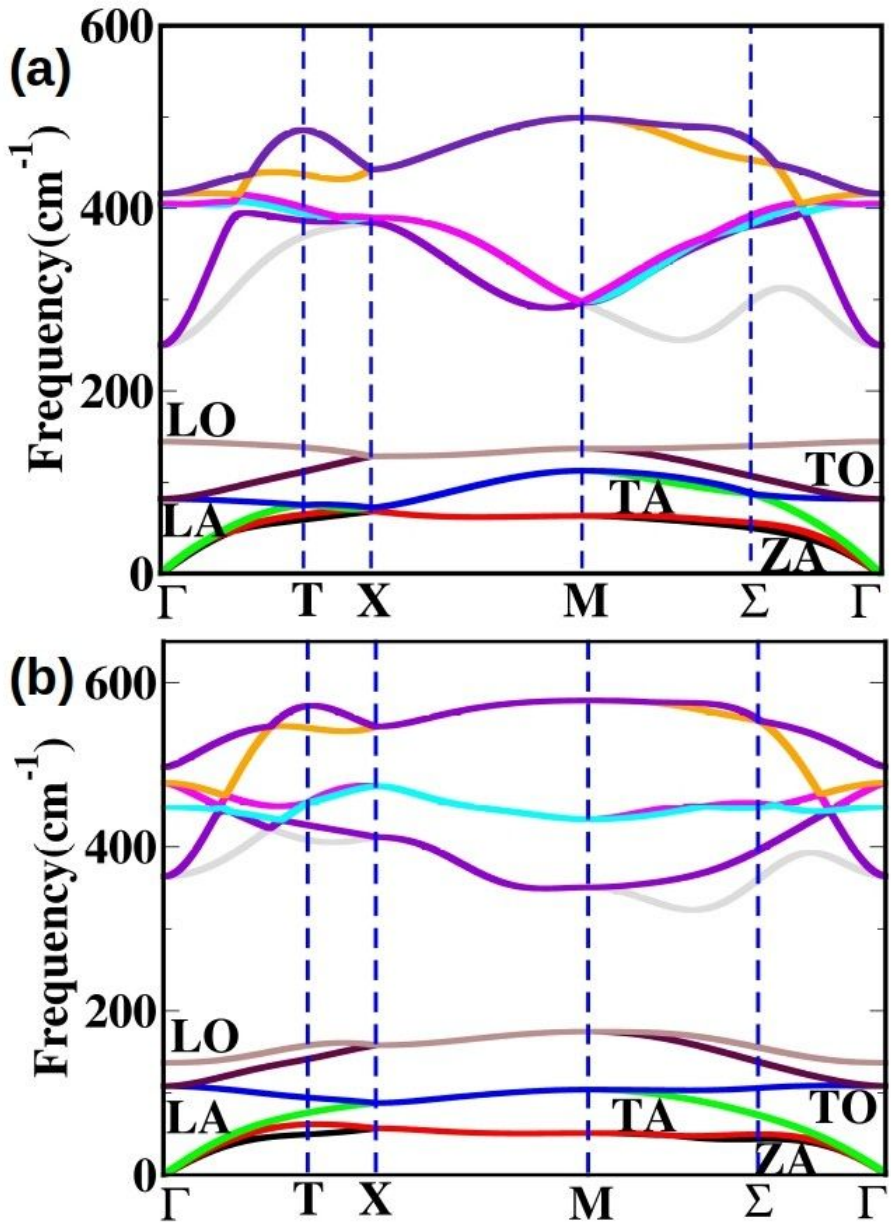

**Figure 3:** Phonon band structure of (a) bulk and (b) monolayer of PbO. Low energy acoustics and optical modes are indicated in the graph using several colours. The bands are represented by different colours like black: ZA, red: TA, green: LA, blue,maroon: TO (transverse optical), gray: LO (longitudinal optical).

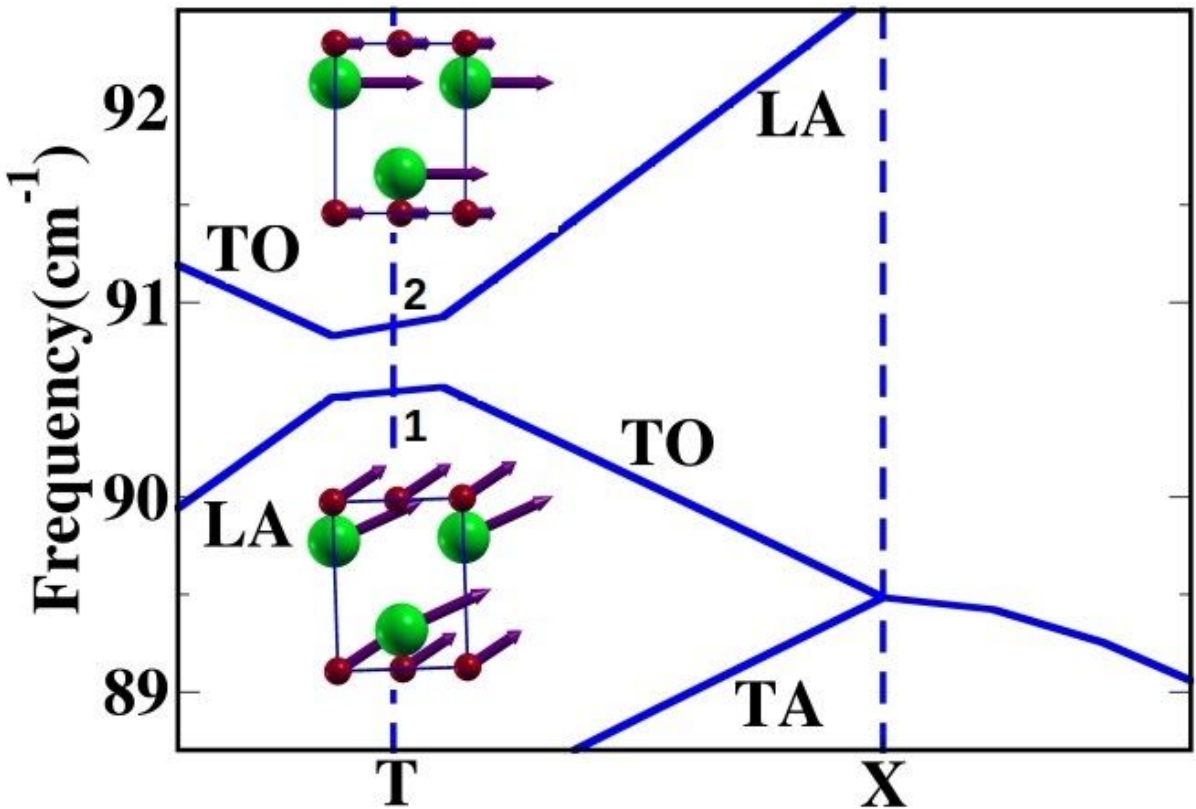

**Figure 4:** Avoided crossing occurred near the high symmetry point X at fine q point. LA and TO bands cross each other at frequency 90.8 cm$^{-1}$. The inset crystal structures signify phonon eigenvectors (represented by arrows) corresponding to bands 1 and 2 at T-point. Both the phonon modes show same point group symmetry ($E_u$) at the vicinity of avoided crossing for SnO and only in bulk PbO.

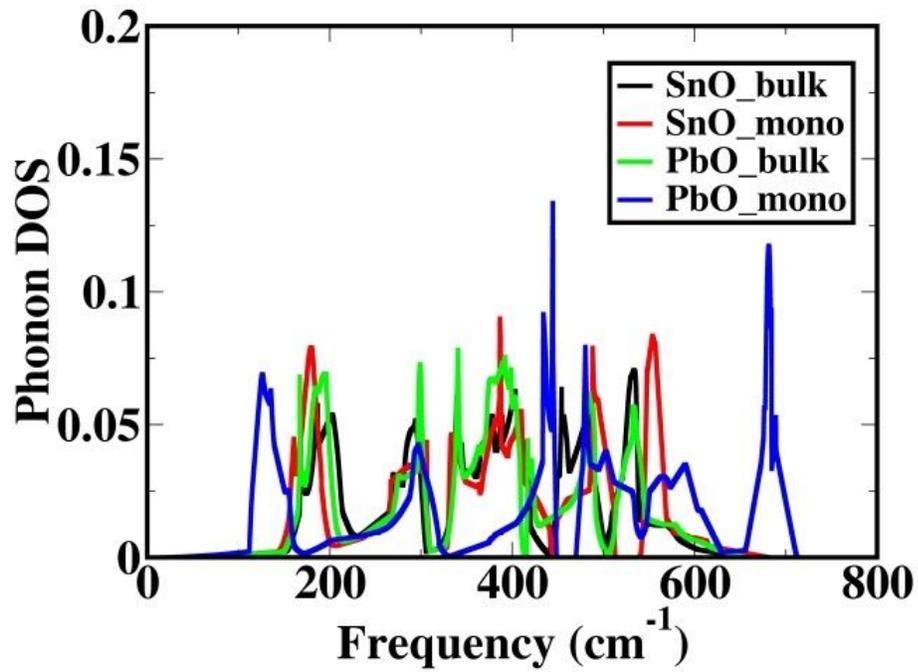

**Figure 5:** Phonon DOS of monolayer PbO only exhibiting significant contribution (blue line) in the low frequency region upto 150 cm$^{-1}$. This is because of the absence of avoided crossing in monolayer PbO.

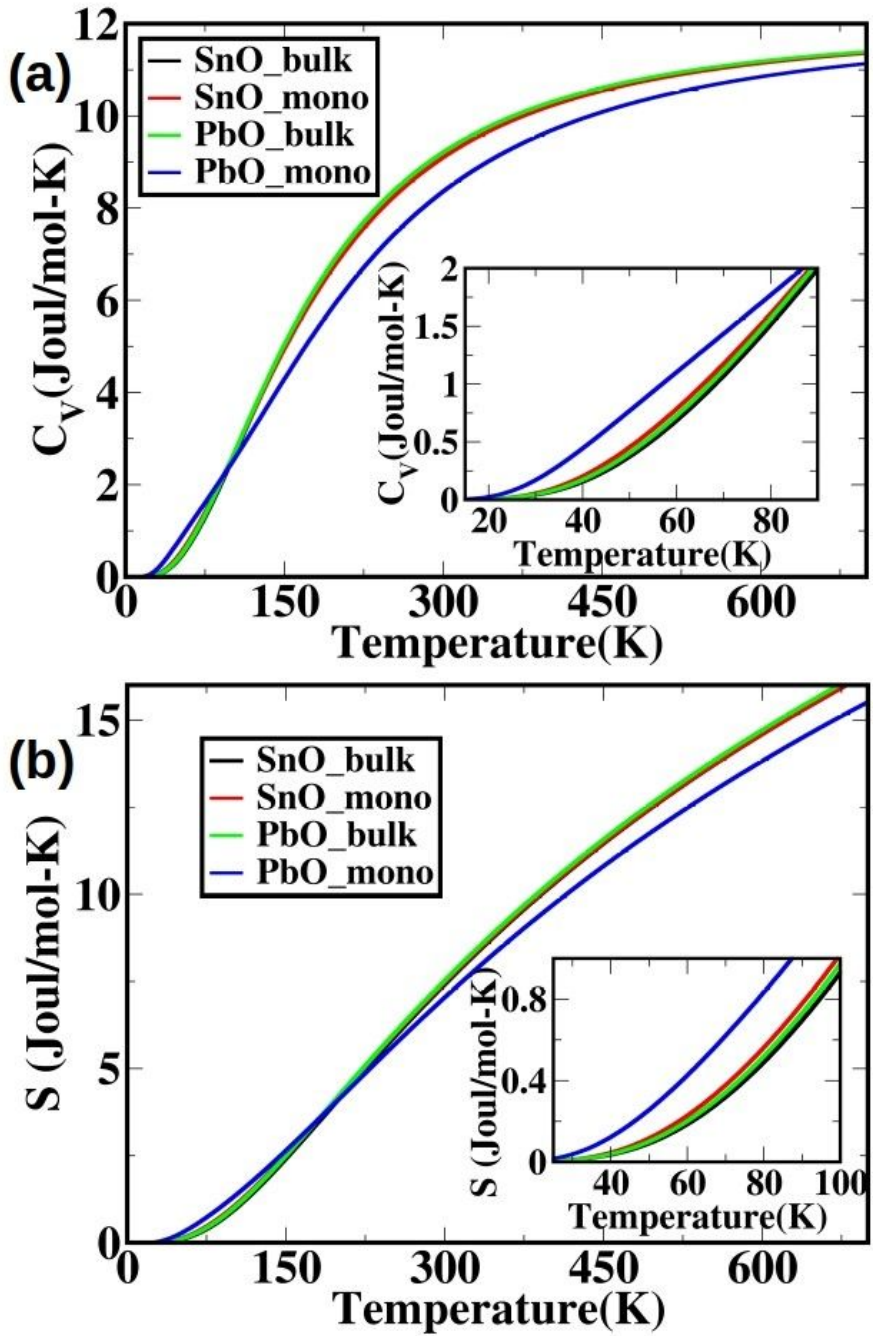

**Figure 6:** (a) Specific heat ($C_v$) and (b) vibrational entropy (S) are plotted with respect to temperature. Specific heat and entropy for single layer PbO is higher than bulk PbO, SnO and single layer of SnO at temperatures up to 150 K. At higher temperature, $C_v$ and S are lower for PbO than SnO because of the heavier mass of Pb. The inset of figure 6a and 6b show specific heat and vibrational entropy upto temperature 100 K respectively (zoomed).

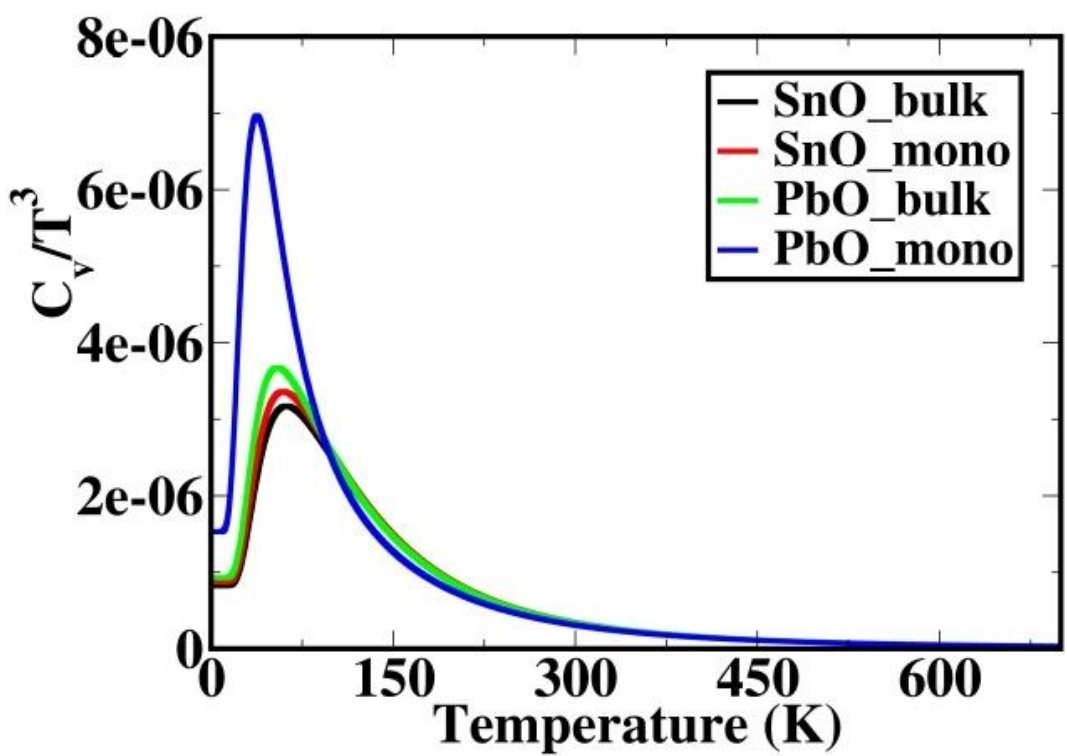

**Figure 7:** Plot of $C_v/T^3$ with respect to temperature (K). Here, $C_v/T^3$ values are given in units of Joul/mol-$K^4$.

**Table I:** Equilibrium lattice parameter (Å) at which DFPT calculation were performed, optical dielectric constant and Born effective charge for the tetragonal phase of bulk and monolayer SnO.

|  | SnO | | | PbO | |
|---|---|---|---|---|---|
|  | Bulk | Monolayer |  | Bulk | Monolayer |
| $a_{cell}$ | 3.87 Å<br>3.80[4] Å, 3.79[6] Å | 3.87 Å<br>3.796[21] Å | $a_{cell}$ | 3.97 Å | 3.97 Å |
| $a_{cell}$ (c-axis) | 5.02 Å<br>4.82[4] Å, 4.81[6] Å | 20.00 Å | $a_{cell}$ (c-axis) | 5.02 Å<br>5.023[9] Å | 20.00 Å |
| $\epsilon_{xx}$ | 7.47 | 2.18 | $\epsilon_{xx}$ | 6.26 | 2.06 |
| $\epsilon_{yy}$ | 7.47 | 2.18 | $\epsilon_{xx}$ | 6.26 | 2.06 |
| $\epsilon_{zz}$ | 6.86 | 1.32 | $\epsilon_{xx}$ | 5.76 | 1.26 |
| $Z^*_{Sn(xy-plane)}$ | 2.98e | 2.69e | $Z^*_{Pb(xy-plane)}$ | 3.08e | 3.26e |
| $Z^*_{Sn(out\ of\ plane)}$ | 2.67e | 0.51e | $Z^*_{Pb(out\ of\ plane)}$ | 2.06e | 0.44e |
| $Z^*_{O(xy-plane)}$ | -3.00e | -2.71e | $Z^*_{O(xy-plane)}$ | -3.09e | -3.27e |
| $Z^*_{O(out\ of\ plane)}$ | -2.71e | -0.54e | $Z^*_{O(out\ of\ plane)}$ | -2.10e | -0.46e |

**Table II:** Phonon frequency (cm$^{-1}$) with optical character is described for bulk and monolayer SnO at zone center position. The crystal plane is along ab-axis and out of plane is along c-axis. Atomic contribution to each mode is also given below.

| SnO-bulk | | | | | SnO-Monolayer | | | |
|---|---|---|---|---|---|---|---|---|
| Frequency | Modes | Character | Atoms | Expt. | Frequency | Modes | Character | Atoms |
| 0.00 | $E_u$ | IR | Sn,O | | 0.0 | $E_u$ | IR | Sn,O |
| 0.00 | $E_u$ | IR | Sn,O | | 0.0 | $E_u$ | IR | Sn,O |
| 0.00 | $A_{2u}$ | IR | Sn,O | | 0.0 | $A_{2u}$ | IR | Sn,O |
| 104.39 | $E_g$ | Raman | Sn,O | | 104.13 | $E_g$ | Raman | Sn,O |
| 104.39 | $E_g$ | Raman | Sn,O | 113[17] | 104.13 | $E_g$ | Raman | Sn,O |
| 199.94 | $A_{1g}$ | Raman | Sn | 211[17] | 199.06 | $A_{1g}$ | Raman | Sn |
| 265.46 | $E_u$ | IR | Sn,O | 260[17] | 275.81 | $E_u$ | IR | Sn,O |
| 265.46 | $E_u$ | IR | Sn,O | | 393.32 | $E_u$ | IR | Sn,O |
| 383.36 | $B_{1g}$ | Raman | Sn,O | | 393.83 | $B_{1g}$ | Raman | O |
| 388.58 | $A_{2u}$ | IR | O | | 482.48 | $E_g$ | Raman | O |
| 458.47 | $E_g$ | Raman | O | | 482.48 | $E_g$ | Raman | O |
| 458.47 | $E_g$ | Raman | O | | 506.95 | $A_{2u}$ | IR | Sn,O |

**Table III:** Phonon frequency (cm$^{-1}$) with optical character is described for bulk and monolayer PbO. The crystal plane is along ab-axis and out of plane is along c-axis. Atomic contribution to each mode is also given below.

| PbO-bulk | | | | | PbO-Monolayer | | | |
|---|---|---|---|---|---|---|---|---|
| Frequency | Modes | Character | Atoms | Expt. | Frequency | Modes | Character | Atoms |
| 0.00 | $E_u$ | IR | Pb,O | | 0.00 | $E_u$ | IR | Pb,O |
| 0.00 | $E_u$ | IR | Pb,O | | 0.00 | $E_u$ | IR | Pb,O |
| 0.00 | $A_{2u}$ | IR | Pb,O | | 0.00 | $A_{2u}$ | IR | Pb,O |
| 81.79 | $E_g$ | Raman | Pb,O | 81[34,35] | 108.25 | $E_g$ | Raman | Pb,O |
| 81.79 | $E_g$ | Raman | Pb,O | | 108.25 | $E_g$ | Raman | Pb,O |
| 144.56 | $A_{1g}$ | Raman | Pb | 145[34,35] | 136.80 | $A_{1g}$ | Raman | Pb |
| 249.99 | $E_u$ | IR | Pb,O | | 363.69 | $E_u$ | IR | Pb,O |
| 249.99 | $E_u$ | IR | Pb,O | | 363.69 | $E_u$ | IR | Pb,O |
| 402.54 | $B_{1g}$ | Raman | O | | 446.00 | $B_{1g}$ | Raman | O |
| 403.66 | $A_{2u}$ | IR | Pb,O | | 477.68 | $E_g$ | Raman | O |
| 415.74 | $E_g$ | Raman | O | | 477.68 | $E_g$ | Raman | O |
| 415.74 | $E_g$ | Raman | O | | 497.10 | $A_{2u}$ | IR | Pb,O |

**Table IV:** Self-force constant for different atoms in the unit cell. B-MO represents bulk MO and M-MO represents monolayer MO in the Table.

| Atom | Direction | B-SnO | M-SnO | B-PbO | M-PbO |
|---|---|---|---|---|---|
| **M** | x/y | 0.144 | 0.144 | 0.143 | 0.269 |
|  | z | 0.225 | 0.253 | 0.244 | 0.277 |
| **O** | x/y | 0.210 | 0.217 | 0.195 | 0.286 |
|  | z | 0.169 | 0.180 | 0.175 | 0.204 |

**Table V:** Interatomic force constants (IFCs) between M-M atoms (M=Sn/Pb) belongs to adjacent layers and M-O (same layer) in the tetragonal unit cell.

| Bonds | Directions | B-SnO | M-SnO | B-PbO | M-PbO |
|---|---|---|---|---|---|
| **M-M** | z | -0.0098 | - | -0.0077 | - |
| **M-O** | x/y | 0.0065 | 0.0067 | 0.0063 | 0.0079 |

Supplemental material for

# Vibrational Spectra of MO (M=Sn/Pb) in Their Bulk and Single Layer Forms: Role of Avoided Crossing in their Thermodynamic Properties

Raju K. Biswas,[1] and Swapan K. Pati[1,2],*

[1]*Theoretical Sciences Unit, Jawaharlal Nehru Centre for Advanced Scientific Research (JNCASR), Jakkur P.O., Bangalore 560064, India*
[2]*School of Advanced Materials and International Centre of Materials Science, Jawaharlal Nehru Centre for Advanced Scientific Research (JNCASR), Jakkur P.O., Bangalore 560064, India.*
*E-mail: pati@jncasr.ac.in*

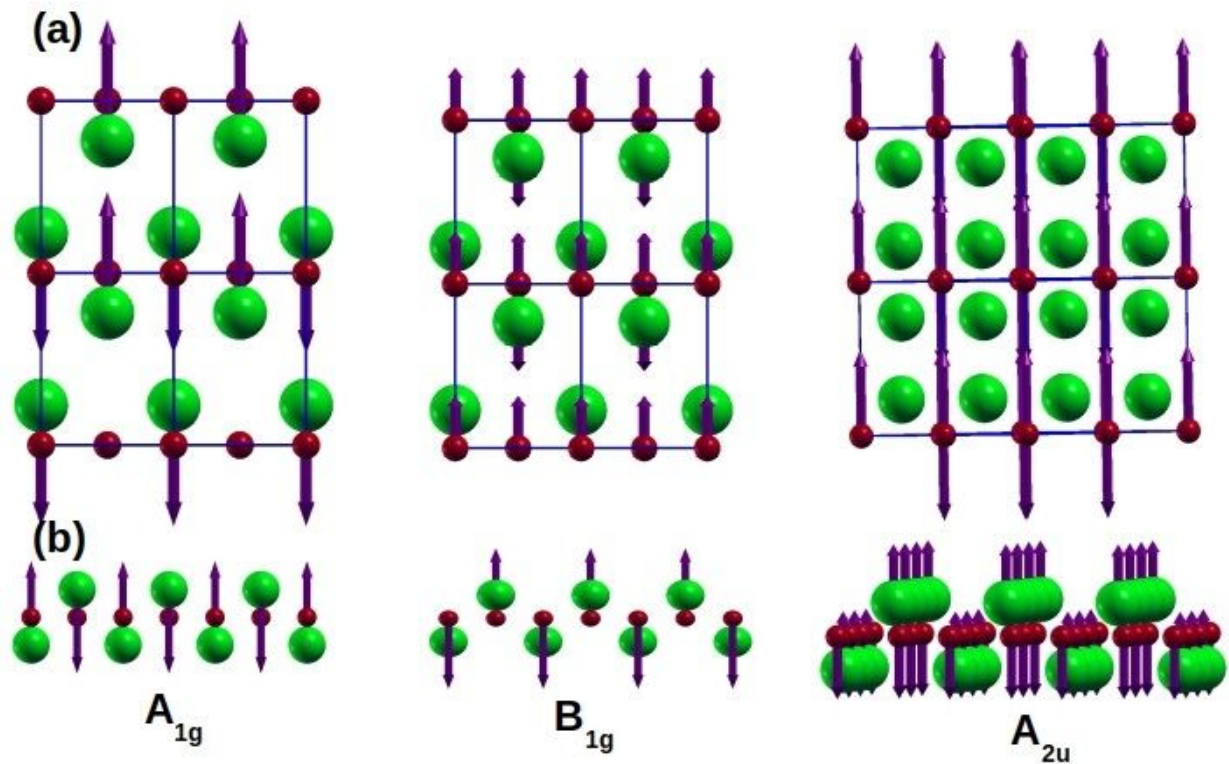

**Figure S1:** Vibrational eigenvectors of (a) bulk SnO, (b) monolayer SnO. $A_{1g}$, $B_{1g}$, $A_{2u}$ are vibrational modes computed at the center of the Brillouin zone ($\Gamma$). Atomic colours are as follows Sn: Green, O: Red. Here, $A_{1g}$, $B_{1g}$, $A_{2u}$ modes are Raman, Raman, IR active respectively. Arrows are indicating direction of forces acting on the atom at stable phase of SnO.

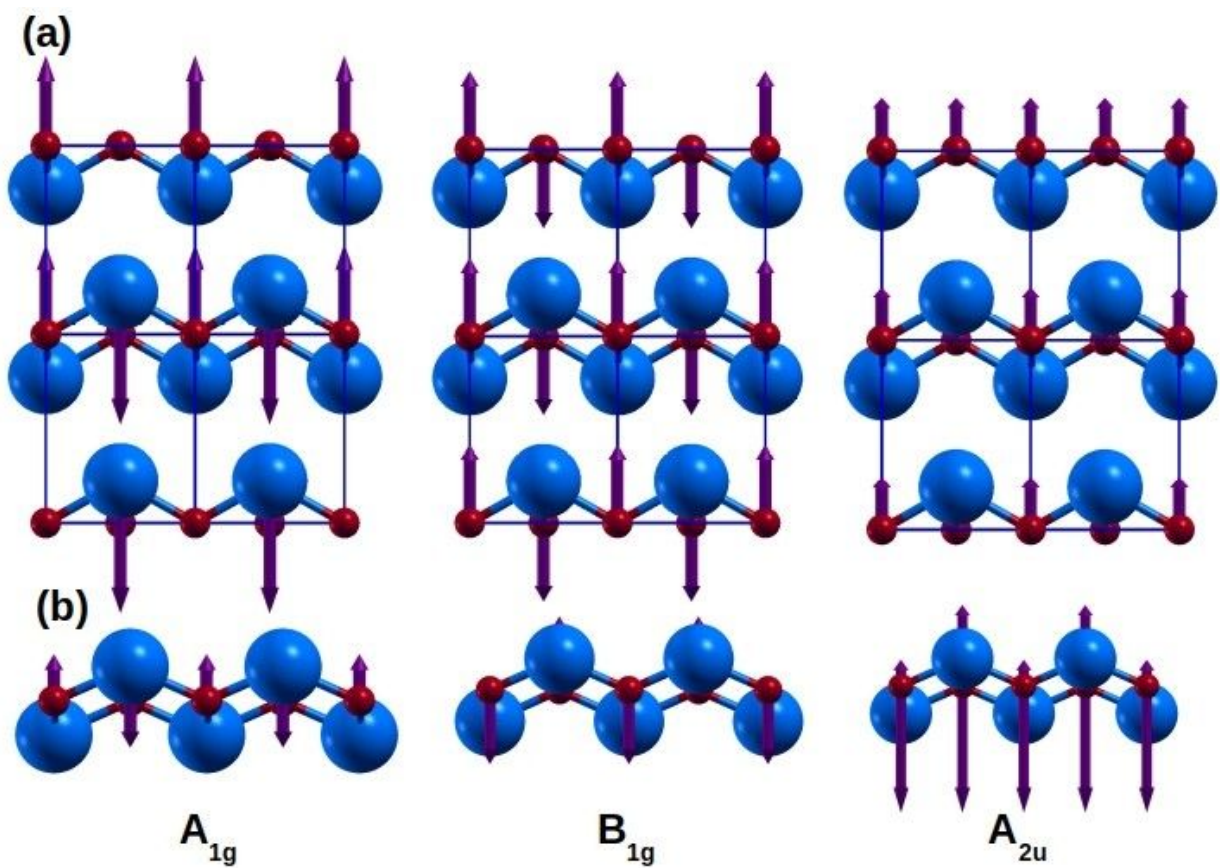

**Figure S2:** Vibrational eigenvectors of (a) bulk PbO, (b) monolayer PbO. $A_{1g}$, $B_{1g}$, $A_{2u}$ are vibrational modes computed at the center of the Brillouin zone ($\Gamma$). Atomic colours are as follows Pb: Blue, O: Red. Here, $A_{1g}$, $B_{1g}$, $A_{2u}$ modes are Raman, Raman, IR active respectively. Arrows are indicating direction of forces acting on the atom at stable phase of PbO.